\documentclass[conference]{IEEEtran}
\usepackage{epsfig,endnotes}
\usepackage{graphicx}
\usepackage{dblfloatfix}
\usepackage{amsmath}
\usepackage{caption}
\usepackage{subcaption}
\usepackage{float}
\usepackage{capt-of}
\begin{document}


\title{SemanticLock: An authentication method for mobile devices using semantically-linked images}
\author{

\IEEEauthorblockN{Ilesanmi Olade}
\IEEEauthorblockA{Xi'an Jiaotong-Liverpool University\\
ilesanmi.olade@xjtlu.edu.cn}
\and
\IEEEauthorblockN{Haining Liang}
\IEEEauthorblockA{Xi'an Jiaotong-Liverpool University\\
haining.liang@xjtlu.edu.cn}
\and
\IEEEauthorblockN{Charles Fleming}
\IEEEauthorblockA{University of Mississippi\\
fleming@olemiss.edu}
}

\IEEEoverridecommandlockouts

\maketitle


\begin{abstract}
We introduce SemanticLock, a simple, fast, and memorable single factor graphical authentication approach for mobile devices. SemanticLock uses a set of graphical images as password tokens that allow constructing a semantically memorable story representing the user's password. Passwords are entered via the familiar and quick action of dragging and positioning user-defined images on the touchscreen. It is well known that for (un)locking mechanisms such as PIN or PATTERN, users tend to pick memorable passwords such as dates or simple (often regular) patterns. This practice by users significantly reduces the effective password space for these mechanisms. The authentication strength of  SemanticLock is based on the large number of possible semantic constructs derived  from the positioning of the image tokens and the type of images selected.  While graphical passwords have been shown in some cases to have lower entropy than other password types, we avoid this problem by (1) performing a series of experiments and analysis to understand which images and image pairs users prefer, and then (2) selecting images that avoid any type of explicit or implicit bias, resulting in an effective password space that is essentially the same as the total password space. Results our study  comparing SemanticLock against other authentication systems show that SemanticLock performs similarly to PIN and PATTERN in usability, while have significantly increased memorability and security.
\end{abstract}
\newcommand{\imgHeightRatio}{0.15}

\section{Introduction} \label{introduction}
Mobile devices, being the de facto personal communication device, are ubiquitous within our society \cite{vonZezschwitz:2013:MGA:2449396.2449432} . We depend on these devices to store substantial amounts of confidential information and perform activities such as emailing, social networking, personal internet banking, and entertainment. All mobile devices manufactured in the last decade come with a default set of authentication or login mechanisms. Research by Micallef et al. \cite{Micallef:2015:WAU:2785830.2785835}, shows that over 64\% of users chose not to secure or use an authentication system on their mobile devices \cite{Harbach}.

While many modern devices use biometric authentication, for example fingerprint or face recognition, to unlock the device, these are vulnerable to spoofing.  Because of this, these devices allow limited authentication attempts via biometrics before falling back to a conventional authentication, usually PIN or Swipe.  In fact, while biometrics is popular for its ease of use, it makes mobile devices \textit{less} secure, because attackers can either spoof the biometric authentication or guess the conventioan authentication option.  On top of this, since the user will infrequently enter the conventional password into their device, remembering this password will be much more difficult, meaning the user is more likely to pick a poor password than if the device did not include a biometric authentication mechanism. This increases the need for mechanisms like SemanticLock, which are easy-to-remember and resist the selection of weak passwords.

In general, research has shown that the behaviour, engagement, and interest of the users have a major impact on the effective  security level of their mobile devices, with many users preferring to sacrifice security for convenience\cite{buschek_snapapp_2016}. The uniformity of distribution  of user passwords within an authentication system's total password space is a practical measure  of the usable level of security of that authentication system. Guessing  or dictionary attacks on user passwords are less successful when authentication systems have a uniform distribution of user passwords. Studies by \cite{Cain:2017:GAR:3027063.3053236,Melicher:2016:UST:2858036.2858384} indicate that the distribution of text passwords chosen by users effectively have very low entropy, meaning that the actual space of passwords most users choose from is much smaller than the total space available. The above observation is known to affect prominent authentication systems such as PIN \cite{Harbach:2016:ASU:2858036.2858267,kovelamudi_scramble_2016,vonZezschwitz:2015:SFS:2702123.2702212} and PATTERN \cite{Zakaria:2011:SSD:2078827.2078835,Uellenbeck:2013:QSG:2508859.2516700,Harbach:2016:ASU:2858036.2858267,vonZezschwitz:2013:PWF:2493190.2493231} and have being extensively studied, with a large body of existing literature.

The PIN authentication system, which is a numeric display of numbers inputted by discrete touches on the screen and the PATTERN authentication system, which is a "grid-like'' display of nodes whose password pattern is selected by a continuous finger movement across the screen to connect the secret password nodes, are both plagued with numerous usage and security issues \cite{Abdelrahman:2017:SCU:3025453.3025461,Mowery:2011:HMC:2028052.2028058,Andriotis:2013:PSS:2462096.2462098,Zakaria:2011:SSD:2078827.2078835}. 
Fortunately, the popularity of touch-screen based mobile devices allows for graphical authentication techniques that offer possibilities of providing passwords that are effectively stronger than text passwords. Recently, researchers have developed and studied various graphical authentication systems \cite{Aly:2016:SGA:2957265.2961863,vonZezschwitz:2013:MGA:2449396.2449432,Davis:2004:UCG:1251375.1251386,Belk:2017:SSU:3106426.3106488,Stobert:2013:MRG:2501604.2501619} that take advantage of the inherent human memorability properties and have attempted to mitigate factors such as low password distribution, low unlocking speed, medium-to-low entropy and other biases, without much success. 

\begin{figure*}[h]
	\centering
	\begin{subfigure}[b]{1\columnwidth}
		\centering
		\includegraphics[width=0.50\columnwidth]{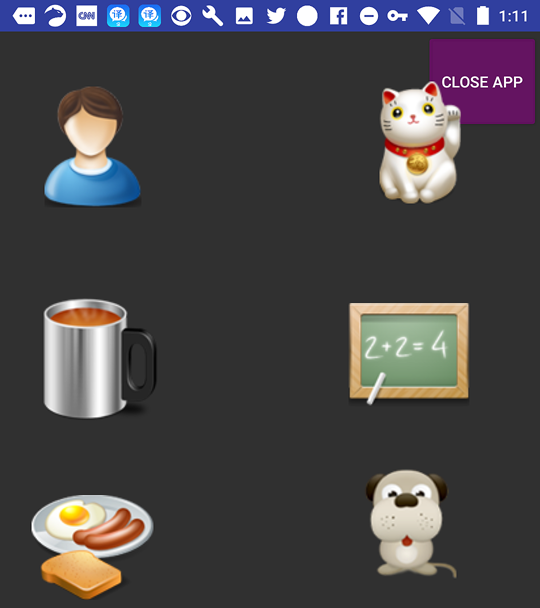}
		\caption{ }
		\label{fig:sub1}
	\end{subfigure}\hfill
	\begin{subfigure}[b]{1\columnwidth}
		\centering
		\includegraphics[width=0.50\columnwidth]{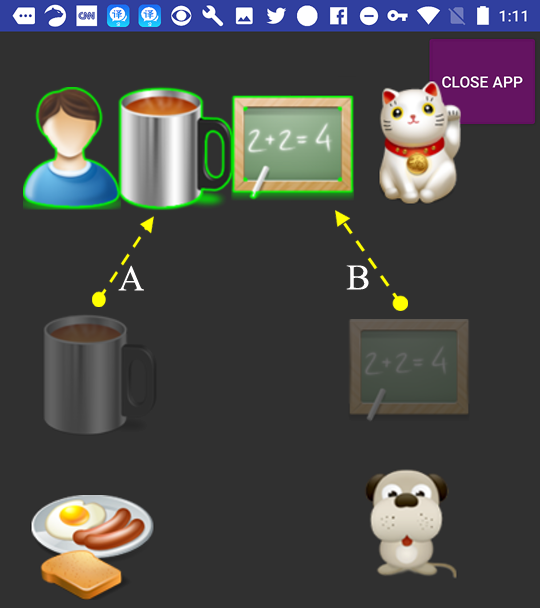}
		\caption{ }
		\label{fig:sub2}
	\end{subfigure}
    	\caption{{\footnotesize \textbf{SemanticLock}: (\subref{fig:sub1}) Default view for login and setup. (\subref{fig:sub2}) Login: the user drags two images to meet a third image. In this case, Cup is dragged to right side of Person \textbf{\textit{(movement ``A'')}}, then Blackboard is dragged to right side of Cup\textbf{\textit{ (movement ``B'')}}. Login can be done with \textit{two quick} movements (\textbf{A,B}).  }}
	\label{fig:Semantic-lockImgUsaged}
\end{figure*}

In this paper, we present SemanticLock, a single factor graphical authentication method for touchscreen mobile devices. Our solution works by providing the user with a way to unlock their mobile devices by joining images via discrete and continuous finger movements to create a semantically memorable story that represents a password (see Figure. \ref{fig:Semantic-lockImgUsaged}(\subref{fig:sub1})). SemanticLock can create a strong memorable password with just two discrete finger movements allowing the user to construct a semantically meaningful password quickly (see Figure. \ref{fig:Semantic-lockImgUsaged}(\subref{fig:sub2})) from the provided images. In the SemanticLock scheme, a password is a sequence of \textit{k} images selected by the user to make a "story'' from a single set of \textit{n $>$ k} images, each non-intrinsically related and placed in position \textit{p}  in one of four locations around a pre-existing image. For mobile devices such as smartphones, six images allows for comfortable usage, yielding 14,400 possible passwords, which is similar to a 4 digit PIN. 

To increase the entropy of the selected password distribution, we reduce password bias by performing a preliminary online study with the goal of eliminating disproportionately popular images and image pairs. In that study, our participants were required to match intrinsically related password images from a set of 40 images that were initially selected from diverse categories (see Figure. \ref{fig:Web_pairboxes}(\subref{fig:Drag_n_Drop_Webpage})). We subsequently obtained 6 \textit{``least intrinsically''} related images from that study and used them during another 2 weeks password creation study (see figure. \ref{fig:web_9by6_page}).\\

In designing the SemanticLock system, we set out to develop a system that was easy to use, very secure and quick to login. Therefore our primary focuses were speed, ease of use, memorability and high entropy. In addition, we wanted our solution to perform consistently across all usage environments and situations our users may find themselves. We ensured that SemanticLock would require only two distinct swipes or finger movements to construct a login password. We implemented a \textit{close proximity} ``sticky'' feature that visually highlights the two images that are in close proximity to each other while the user is actively dragging one of the images. If the user releases this image it automatically ``glides'' towards the closest image  and ``sticks'' to it. This feature greatly reduces errors caused by unsteady finger movements and increases overall login speeds. SemanticLock also inherits both the discrete and continuous finger movement properties of the PIN and PATTERN authentication system respectively. However, in contrast to PATTERN, SemanticLock only requires two short swipes rather than one continuous long swipe thereby minimizing the time needed to complete a login session or recover from errors \cite{Riva:2012:PAD:2362793.2362808}.

Through a series of studies, we evaluate SemanticLock from usability, memorability, and security, perspectives, comparing it against the two most commonly used mobile password systems: PIN and Swipe.  We show that SemanticLock is similar in usability to both of these methods, but superior in terms of security and memorability.

\section{Related Work on Authentication Methods}\label{Related_work_on_Authen}
User authentication and access control  are very important in today's electronic world. The advent of personal computing and mobile devices has made security a foremost consideration in the design and usage of these devices. While authentication can exist in many forms, there are three core types of authentication categories with which a user can be identified by a system. These categories are namely: \textit{What you know, What you have} and \textit{What you are}. The practical implementation of these categories are the text , graphical passwords, token based passwords, and biometric passwords. We next examine the history and various studies pertaining to text and graphical password implementations.

\subsection{Text Based Passwords}
Alphanumeric text-based passwords have dominated human-computer authentication since the 1960s \cite{DeLuca:2010:CSP:1753326.1753490}, where keyboards were used to input user passwords. With the emergence of mobile devices with 10 digit keypads \cite{vonZezschwitz:2015:SFS:2702123.2702212}, the use of numeric-based PIN passwords became mainstream. The first generation touch-screen based smart-phones featured numerous variants of PIN-based password systems \cite{Chiang:2013:IUA:2493190.2493213,Haque:2013:PIT:2516760.2516767}, has been used by all mobile device form factors \cite{Aly:2016:SGA:2957265.2961863,Cain:2017:GAR:3027063.3053236,Dunphy:2010:CLR:1837110.1837114} and remains very popular with users. Although the text-based and the PIN passwords have high theoretical password spaces, numerous studies, such as those by Bonneau et al. \cite{10.1007/978-3-642-32946-3_3} and  Melicher et al.\cite{Melicher:2016:UST:2858036.2858384} show that the practical password spaces and entropy are very low due to user security behaviours. For many years the security literature lacked sound methodology and ecological validity \cite{Fahl:2013:EVP:2501604.2501617} to answer elementary questions about the practical password distribution, or the effects of demographic properties on their outcome. Consequently, there remained an open question as to the extent to which passwords are weak due to a lack of motivation or inherent user limitations \cite{Bonneau:2012:SMI:2437647.2437657,Bonneau:2015:PEI:2797100.2699390}. In a study by \cite{10.1007/978-3-642-32946-3_3}, it was shown, based on an available large public dataset of PINs, that 29\% of the selected 4-PIN and 6-PIN passwords correspond to a date based sequence. This significantly reduces the practical password space of PIN passwords. The massive disclosure of millions of real-life user passwords in hacked password databases \cite{6234435,Jakobsson:2012:BUP:2372387.2372397,Veras:2012:VSP:2379690.2379702} from several websites such as RockYou, Yahoo, Hotmail, Flirtlife and Computerbits, exposed an enormous gap between the real password distribution and the theoretical space of passwords. Furthermore, analyses by Malone et al. \cite{Malone:2012:IDP:2187836.2187878} observed that security motivations such as registering payment cards or supplying sensitive financial information did not affect the users tendency to create weak passwords. In the final analysis, practical user passwords distribution is skewed towards low password entropy and protection. Additionally, studies by Melicher et al. \cite{Melicher:2016:UST:2858036.2858384} confirm that this pattern of skewed password distribution and low password entropy is worse with mobile device users due to additional restrictive factors inherent with using mobile devices, such as limited screen size, restricted hand and finger access to the entire keypad \cite{Ng:2014:IEE:2556288.2557312,Ng:2015:EEM:2785830.2785853,Feng:2015:IPI:2786567.2793711}.

\par
\subsection{Graphical Passwords}
A graphical password, a term introduced by Blonder \cite{BlonderG69}, is an authentication system that is presented to the user via a graphical user interface (GUI), and from a smart mobile device perspective, this GUI includes a touch-screen system that enables easy interaction with the objects displayed on the GUI. Graphical passwords provide a promising alternative to traditional alphanumeric passwords. They are attractive and intuitive since people usually remember shapes and images better than random words or text. In recent years, various studies have categorized graphical authentication methods into 3 main categories:

\textbf{Recall:} The \textit{Recall} graphical authentication system gets its origin from works done on Draw-a-Secret \cite{Zakaria:2011:SSD:2078827.2078835}, Pass-Go \cite{Tao2008PassGoAP} and other similar systems. It is shown to be a memory intensive task \cite{Biddle:2012:GPL:2333112.2333114} due to the fact that the secret diagram or pattern initially drawn by the user has to be entirely remembered and reproduced. The advantage of \textit{Recall} is that it benefits from the inherent motor memory of the users and our superior ability to recall shapes and patterns \cite{Dunphy:2010:CLR:1837110.1837114,Uellenbeck:2013:QSG:2508859.2516700}. The Android Pattern password system is recall-based.

\textbf{Recognition:} The recognition graphical authentication systems revolve around the ability of the user to \textit{'recognize'} sets of images from among decoys that had been selected earlier during the initial creation of the passwords. Recognition based systems such as Passfaces \cite{Passfaces1,10.1007/978-1-4471-0515-2_27} , D\'{e}j\`{a} vu \cite{Dhamija:2000:DVU:1251306.1251310} have been extensively studied. An advantage of this system in certain implementations, such as Passfaces, is our intrinsic ability to recognize human faces, while on the other hand, this ability induces biases in our selection of these faces to start with, and it is also a memory intensive process. This was observed in a study by Davis et. al \cite{Davis:2004:UCG:1251375.1251386}, where users often choose faces of members of their own race.

\textbf{Cued-recall:} Cued-recall based systems exploit various studies that conclude that the human memory holds information that may be available yet inaccessible for retrieval without the proper trigger or catalyst \cite{Chiasson:2007:GPA:2393847.2393880}. This system is based on the idea that pictorial indicators can simplify the task of recall for a user \cite{Uellenbeck:2013:QSG:2508859.2516700}. A major disadvantage of these graphical cues is that they may constitute inadvertent \textit{``hotspots''} that also serves to weaken the password strengthen of the authentication system. Cued-recall based systems such as PassPoint \cite{WIEDENBECK2005102} and  Cue Click Points(CCP) \cite{Chiasson:2009:UID:1667545.1667546} have been extensively studied.

\begin{figure*}[ht]
	\centering
	\begin{subfigure}{1.3\columnwidth}
		\centering
		\includegraphics[height=0.2\paperheight]{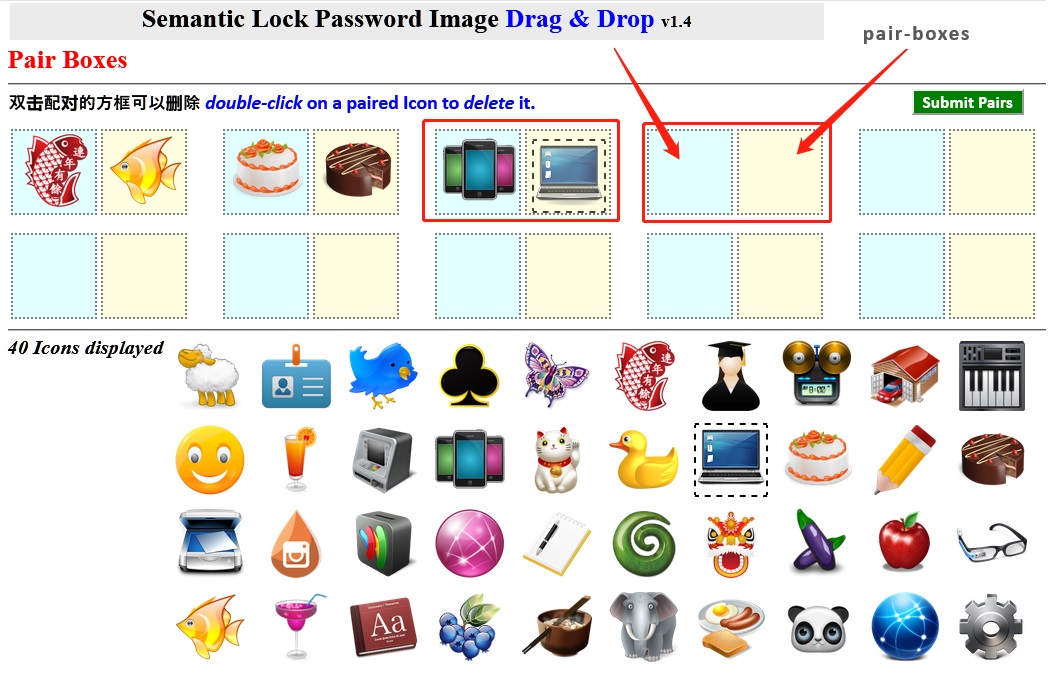}
		\caption{ }
		\label{fig:Drag_n_Drop_Webpage}
	\end{subfigure}\hfill
	\begin{subfigure}{0.7\columnwidth}
		\centering
		\includegraphics[height=0.2\paperheight]{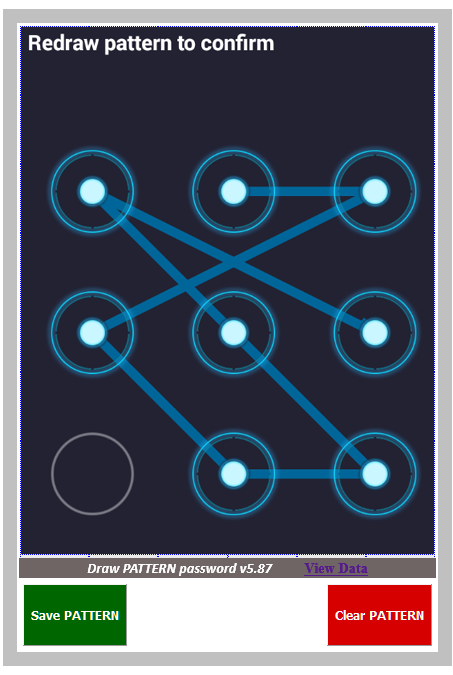}
		\caption{ }
		\label{fig:WebPage_Pattern_Interface}
	\end{subfigure}
	\caption{{\footnotesize \textbf{Data Collection Web-Pages}: (a) \textit{Related Icon Pairing Web Interface:} Our online web page allowed our participants to select 2 icons that they felt were related. They dragged these icons into the ''pairboxes``.\textbf{(b)} \textit{PATTERN password Web Interface :} Participants were requested to create various pattern passwords.}}
	\label{fig:Web_pairboxes}
\end{figure*} 
 As stated above, it is obvious that graphical password authentication systems exist in different implementations, and have been extensively studied. A study by von Zezschwitz et al. \cite{vonZezschwitz:2013:MGA:2449396.2449432} compared three custom graphical authentication systems against the PATTERN authentication system. The aim was to study their prototypes' unlock speed, level of memorability, usability and user acceptance. Results confirmed that PATTERN authentication system was superior to their proposed prototypes in regards to unlock speed and performed comparatively similar in regards to usability, user acceptance and memorability but was considered less secure by the users. It should be noted that the PIN authentication system was not included in their study, and also that the  effective password distribution or password space was not evaluated in this study either. In a later study,  von Zezschwitz et al \cite{vonZezschwitz:2013:PWF:2493190.2493231} compared the PIN and PATTERN authentication systems, and the results indicated that PIN had a faster unlock speed and smaller error rate, but the PATTERN was more usable, memorable and likeable. However, studies of user Pattern password creation by \cite{Uellenbeck:2013:QSG:2508859.2516700,Aviv:2015:BBC:2818000.2818014,Tupsamudre:2017:PPI:3052973.3053041}, found empirically that there is a high bias in the Pattern selection process resulting in low entropy and a practical effective security of less than a three digit randomly-assigned PIN.

More recently Aly et al. \cite{Aly:2016:SGA:2957265.2961863} introduced SpinLock, a technique that is based on a physical combination lock, and require users to rotate a dial both counter-clockwise and clockwise alternatively to select a password token. This design is meant to make it usable but without sacrificing security. Their study with 21 participants using SpinLock in 63 trials with various degrees of password complexity show that it could lead to significantly lower time performance than Pattern Lock and only achieved similar performance with PIN. Their participants thought that SpinLock was more usable and enjoyable to use.

\section{Methodology}\label{Methodology}
Our experiments for SemanticLock were split into two parts.  The first part was to enable the selection of an icon set with minimal single icon or pairwise bias.  This portion of our experiments was web-based and is outline in Section \ref{webStudy}. The second part was to evaluate the performance of SemanticLock as a mobile authentication system.  The evaluation part of our experiments was split into two parts: usability and memorability comparison of SemanticLock vs PIN and Swipe, and a long-term, daily use, usability study of SemanticLock.  

We split the evaluation in this way for practical reasons.  First, if the participants are using SemanticLock as their primary locking mechanism for their mobile device, memorization rates for their passwords will be 100\%, making it impossible to compare the relative memorability of the three systems.  Second, all of the users in the long-term, daily use portion of the study had used both Swipe and PIN systems in the past, so we did not feel it was necessary to run parallel control groups with these two methods for this phase of the study. Instead for our daily use study, we chose Swipe users as half of our participants and PIN users for the other half, allowing use to compare against the long term usability of these systems.

\subsection{Participant recruitment and ethical concerns}
Because our study utilizes human subjects, it was reviewed and approved by our University Research Ethics committee.  One concern raised was whether or not participants would inadvertently reveal their personal passwords.  To prevent this, a text was added in the instructions of each experiment specifically warning participants not to use any current or past personal password as part of their responses.

Participants for all experiments were volunteers, recruited via a university-wide e-mail. Participants were primarily students, but also included some staff members.  All participants were daily mobile device users. Participants in the web survey portion of our study were not compensated, due to the number of participants and the relatively short duration of their participation.  Volunteers for the longer term memorability and usability studies were compensated with coffee shop gift cards worth between 5 and 30 USD, depending on the duration of participation.

\subsection{Web Study}
For this aspect of the  study, we utilized multiple web-based interfaces that were designed using HTML5, PHP and MySQL database back-end technologies. This allowed us to implement icon drag-n-drop actions and graphical line drawing functions that are common on touch-screen based devices. This web-based approach allowed us to collect  large amounts of data from our participants at various locations and use this data for preliminary determination of icon selection for SemanticLock and practical password entropy evaluations \textit{(see section \ref{Authen_Sec_Entropy})}  for both the PATTERN and SemanticLock authentication system (see figure \ref{fig:Web_pairboxes}). Although web-based experiments are harder to control than in a laboratory or supervised field experiments \cite{Biddle:2012:GPL:2333112.2333114}, this channel of data collection meets our requirements and offers numerous advantages.

\subsubsection{Goals}
As part of our goals in the design of our SemanticLock system, our initial intention is to avoid any implicitly induced biases in the researcher's selection of the password icons that may lower the entropy or reduce the achievable password space \cite{DeLuca:2010:CSP:1753326.1753490}. In general, security experts have observed that an authentication system's \textit{theoretical password space} is never optimally achieved during practical usage \cite{Cha:2017:BGA:3052973.3052989,Haque:2013:PIT:2516760.2516767,6234435,Melicher:2016:UST:2858036.2858384}, and there is also a need to determine the actual \textit{practical password space} that supports the ecological validity of such an authentication system.  We defined two stages of experiment to achieve the above stated objectives, and implemented these stages with two different groups of participants. The output of the analysis of the dataset collected in the first stage was utilized during the second stage. Our goal for also collecting PATTERN password data was to acquire data to be used to determine the password strength and other result comparisons (see Section \ref{Authen_Sec_Entropy}).
\subsubsection{Participants}
\paragraph{For Stage 1:} We engaged 372 participants, mostly university students, but with diverse age ranges. Our participant group included 45\% female users; we also collected other demographic information such as academic background, computer skills and their experience with mobile devices or authentication systems.
\paragraph{For Stage 2:} We engaged 184 participants, 70\% were students within the same university campus and the rest were non-students. Our web portal included a 3 minute training video, and each participant was encouraged to watch the video before attempting to create passwords. We advised our participants to create at least 10 passwords each. Our participant group included 48\% female users; we also collected other demographic information such as academic background, computer skills and experience with mobile devices or authentication systems.
\subsubsection{Acquisition of independent password icons} \label{Experiment_Design_stage1}

Our initial process was to provide a set of 40 icons that were drawn from various categories and genres. We explicitly avoided icons that had major gender oriented colors, and icons with cultural, national or religious relevance. Our participants were then presented with a web-based interface that displayed these icons on a 10 x 4 grid, with each icon randomly positioned in different grid-cells during every selection session to prevent locational bias. Participants were required to create 10 sets of \textit{"icons-pairs"} that they believed were related by dragging these icons into the provided \textit{`pairboxes`} (see figure. \ref{fig:Web_pairboxes} (\subref{fig:Drag_n_Drop_Webpage})), the reason or logic of this relationship was based on their discretion. Each participant was allowed multiple iterations. We analyzed the 3708 collected \textit{pair-datasets} to extract 6 icons that were the least intrinsically related. These \textit{"non-intrinsically"} related icons were used in the next stage of the experiment. Secondly, the participants were also shown a 3x3 PATTERN web interface (see figure. \ref{fig:Web_pairboxes} (\subref{fig:WebPage_Pattern_Interface})) and were requested to draw 10 different patterns. The web-interface ensured that the user could not repeat patterns within the same session or create patterns with less than 3 nodes.
\begin{figure*}[t]
	\centering
	\begin{subfigure}{1\columnwidth}
		\centering
		\includegraphics[height=0.15\paperheight]{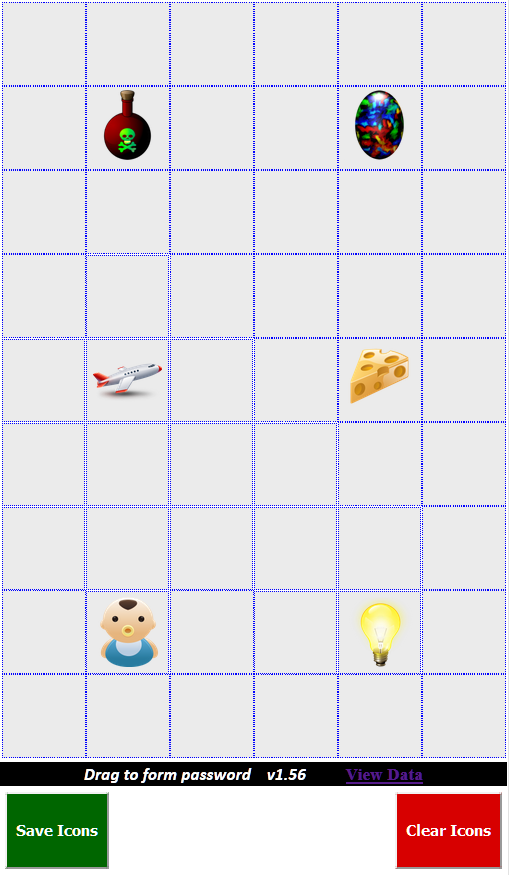}
		\caption{ }
		\label{fig:pwd1}
	\end{subfigure}
	\begin{subfigure}{1\columnwidth}
		\centering
		\includegraphics[height=0.15\paperheight]{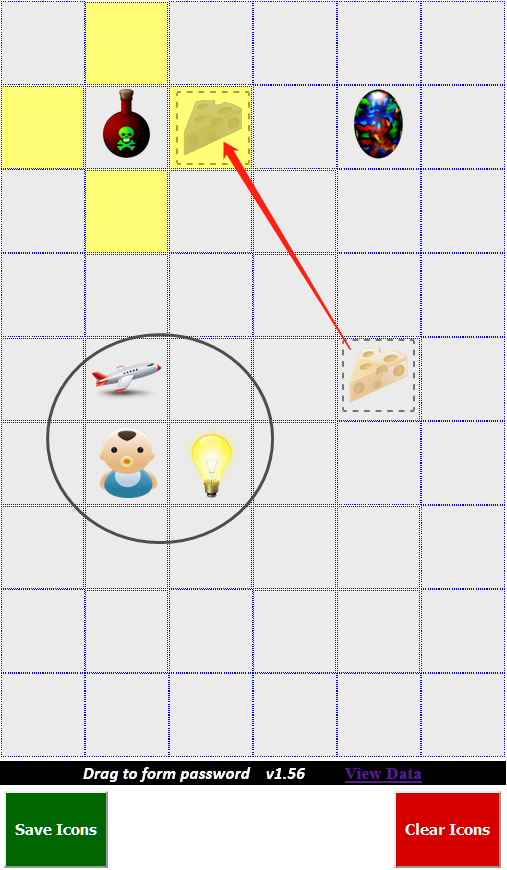}
		\caption{ }
		\label{fig:pwd2}
	\end{subfigure}
	\caption{ {\footnotesize \textbf{SemanticLock Web-based Password Creator}: (\subref{fig:pwd1}) Default view of icon placement. (\subref{fig:pwd2}) \textbf{Creating Password:} the user drags the ``cheese'' to meet the stationary ``bottle'' icon. In this case, ``cheese'' is also dragged to right side of ``bottle''. Lastly a three-icon password is shown \textbf{\textit{ (see black circle).}} } }
	\label{fig:web_9by6_page}
\end{figure*} 
\subsubsection{Data collection for the evaluation of practical password space} \label{expstg2}
Our primary goal was to quantify the effect of a participant's choice on the security of passwords chosen. Every authentication scheme has an entropy and the strength of such entropy is determined by the probability distribution associated with the password space \textit{(see section \ref{Authen_Sec_Entropy})}.  Ideally, this distribution is approximately uniform. At this stage of our experiment we presented a SemanticLock web-based interface displaying the six derived \textit{non-intrinsically related}  password icons on a 9 x 6 celled grid to our participants (see figure. \ref{fig:web_9by6_page}(\subref{fig:pwd1}) ). Our participants were required to create several semantic passwords with the password icons by dragging a chosen icon to the \textit{left, top, right or bottom} position of an associated stationary icon (see figure. \ref{fig:web_9by6_page}(\subref{fig:pwd2})). During this stage we also presented the users with 3x3 Pattern web-interface (see figure. \ref{fig:Web_pairboxes} (\subref{fig:WebPage_Pattern_Interface})) and requested that they create 10 unique passwords that has more than 3 nodes. We succeeded in collecting password data for both the SemanticLock and PATTERN authentication system that would be used to evaluate the practical password strength of these systems.

\subsection{Mobile Device Memorability and Comparative Usability Study} \label{mobdevstudy}
Our mobile device study made use of the Android platform. We developed a mobile version (see Figure. \ref{fig:Semantic-lockImgUsaged}) of the interface that was used during our web-based study (see Figure. \ref{fig:web_9by6_page}). We also developed Android versions of the Pattern and PIN lock authentication systems since these authentication systems would be our baseline or control conditions for this study due to their popularity and  large body of research literature  about their performances. We developed an additional application to help us convey the testing and survey to our participants in a uniform and consistent way. It allowed participants to view an initial training video, assigned a unique participant ID that allowed us to correlate data across login techniques and also presented the pre-survey and post-survey questionnaires in the proper sequences while implementing the Latin square approach to counterbalance the order of the techniques.
\subsubsection{Goals}
Our goal during this study, which involved participants in an indoor environment, was to collect both qualitative and quantitative data which would provide  insight into our participant's perception of the likeability, usability, memorability and login speed of the 3 authentication systems: SemanticLock, Pattern Lock, and PIN.    

The PIN, PATTERN and SemanticLock applications developed for this study meet our goal of ensuring compatibility with Android 6.0 and above, while meeting the requirements of working on phone and tablet mobile device form factors. The training mode option allowed users to receive adequate training and practice before the actual testing. During the testing, a participant's activities such as touches, password tokens, strokes, pauses, timings, aborts and errors were logged for further analysis.
\subsubsection{Participants}

We recruited 63 participants (35 females) for this phase of our study. The data from our pre-testing survey reveals that 51\% of the participants were between the ages of 17 to 27 and all our participants were right-handed. All were active users of iPhone (31\%) and Android (66\%) mobile phones. 55\% of them used a phone with fingerprint sensor, while 17\% used the PIN, 14\% used PATTERN, and the remaining 14\% did not use authentication. 

\subsubsection{Experimental Design}
Our goal was to compare the three main techniques and their interactions with other independent variables. To do this, we followed a within-participants design. The independent variables in our study are: Technique  \textbf{\textit{{\footnotesize (PIN, PATTERN, SemanticLock)}}} and Device form-factor   \textbf{\textit{{\footnotesize (Phone, Tablet)}}}. The dependent variables are: login speed, pre-login delay time, error rate, user usability rating, user acceptance rating, and user perceived speed rating.

\paragraph{\textbf{Technique:}}
Our experiment compared three techniques which are the PIN, PATTERN and SemanticLock authentication systems. The task required of each participant was to enter the password tokens as fast as possible during each session, whereby we implicitly collected and tracked data and meta-data for further analysis. We assigned password tokens for each technique so that each participant would use a sufficiently strong password properly distributed within the space of possible passwords. We attempted to ensure that the password tokens given for each technique had relatively the same password strength.  

The decision to assign passwords was based on an  exploratory experiment we performed as part of the experimental design.  In these initial experiments the majority of users selected extremely weak passwords, such as ``1111'' or ``1234''. In our discussions with participants, we found that when asked to create a password with no risk of data loss, they opted to choose the simplest acceptable password possible, despite all agreeing that they would not use this type of password on their personal devices.  Since these types of passwords would not result in meaningful results, we opted to all participants to choose from a set of pre-generated passwords of similar entropy.

\paragraph{\textbf{Device Form-Factor:}}
Mobile devices are available in various dimensions. We performed our study with a 5.2'' LG Nexus 5X phone and a 10.2'' Google Pixel C tablet. 

\subsubsection{Task and Procedures}
Our first step was to inform the participants about the confidentiality of their supplied information and to explain the purpose of the project and the tasks they would need to do. We provided a three minute training video to each participant, after which they were allowed to practice each technique a couple of times. They practiced the creation of a password and the use the password to login into the mobile device. We emphasized the need for a speedy and accurate login during the actual testing phase.

	\textbf{Week 1 (First Phase):} 
    Each participant was required to answer a pre-test questionnaire before commencing the test. We allowed each participant to choose password tokens for each technique from our supplied list. If the participant entered a wrong password, the application alerted them to enter the correct password again. The average time for participants to complete all techniques (including questionnaires) was 4 minutes. The experiment finished with a Likert questionnaire that collected qualitative data about the participants' perceived usability, error-handling, security and likeability of each technique. The participants used the techniques on the LG mobile phone and the Google tablet. The main independent variables were \textit{technique} (PIN, Pattern and SemanticLock) and\textit{ mobile form factor} (phone and tablet). Each participant had to enter a total of 9 passwords per session, 3 for each Technique and participants were allowed a 60 second rest in between techniques to minimize fatigue, if there was any.
    
    \textbf{Week 2 (Second Phase): } 
    In the second phase, which occurred a week after, we explored the memorability aspects  of the three techniques. We asked the same participants to recall the passwords they had used for each technique the previous week. During this session we tracked error-rates, type of error, action-delay times and login speed required for our future analysis.

\subsection{Long-term Usability Study}
In order to evaluate the suitability of SemanticLock as a replacement for PIN or Swipe, we recruited 10 volunteers (5 male and 5 female) to use SemanticLock as the primary lock mechanism on their mobile device.  Half of these volunteers were PIN users and the other half were Swipe users.  At the end of two weeks, we gave the participants a follow-up questionnaire to determine their thoughts on SemanticLock.  


\section{Data Collection \& Measurement} \label{Data_Collection}
We collected data for a number of dependent variables and used this data to evaluate the techniques.

\textbf{Pre-login delay} is the elapsed time between when the participant indicated that they were ready to start unlocking the device and the actual time they entered the password. This data provides a view into evaluating the memorability and usability of the system. Studies by Stobert et al. \cite{Stobert:2013:MRG:2501604.2501619,Weiss:2008:PUS:1463160.1463202} defined a direct relationship between memorability and pre-login delay time. We analyze this data to quantify the level of memorability and usability.

The time period used to complete each trial of the login process for a technique was recorded. This measurement only recorded successful trials; failed trials were recorded as singular failure events. Login speed was tracked from the moment a participant starts password token entry until the entry was completed successfully.

The error rate was measured as a percentage of failed login attempts to the total number of attempts required to complete the technique's session. The number of failed login attempts during a trial did not affect the number of trials that constituted a complete session. 

We collected \textit{pre-test}, \textit{in-test} and \textit{post-test} surveys via an electronic questionnaire. The questions focused on ease of use, perception of speed, likelihood of adoption, error recovery, and interface usability. We implemented the questionnaire in electronic form and used 5-point Likert questions for some aspects of the questionnaire.

\section{Results} \label{Results}
\begin{figure*}[ht]
	\centering
	\begin{subfigure}[b]{.22\textwidth}
		\centering
		\includegraphics[height=0.15\paperheight]{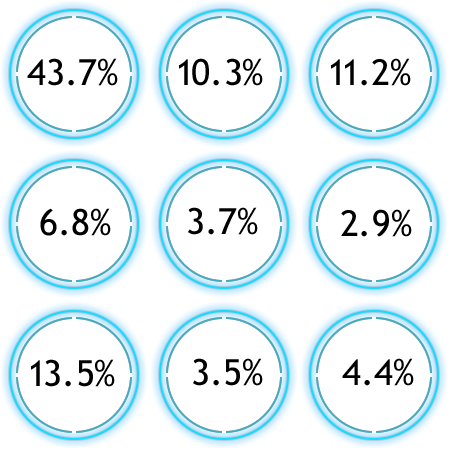}
		\caption{{\scriptsize Pattern Start points}}
		\label{fig:3_x_3_StartPoint}
	\end{subfigure}\hfill
	\begin{subfigure}[b]{.22\textwidth}
		\centering
		\includegraphics[height=0.15\paperheight]{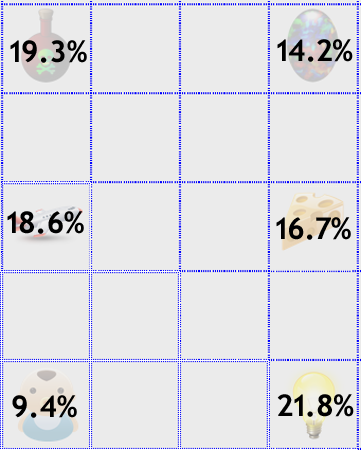}
		\caption{{\scriptsize SemanticLock Start points}}
		\label{fig:Web_study_password_StartPointa}
	\end{subfigure}\hfill
	\begin{subfigure}[b]{.22\textwidth}
		\centering
		\includegraphics[height=0.15\paperheight]{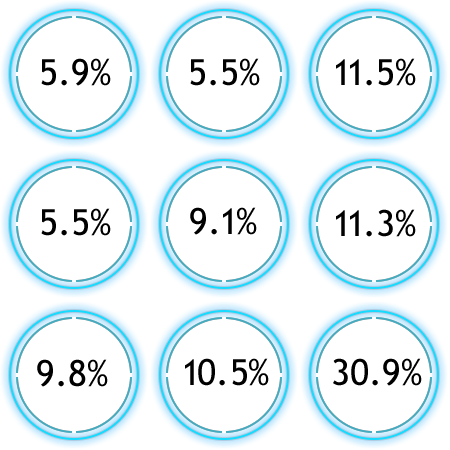}
		\caption{{\scriptsize Pattern End points}}
		\label{fig:3_x_3_endPoint}
	\end{subfigure}\hfill
	\begin{subfigure}[b]{.22\textwidth}
		\centering
		\includegraphics[height=0.15\paperheight]{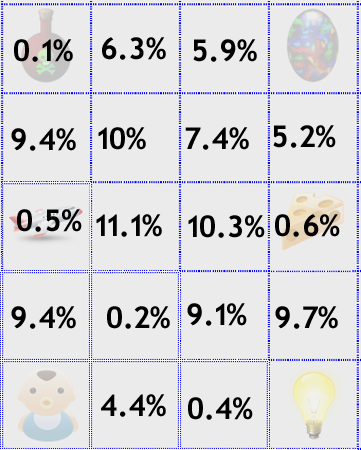}
		\caption{{\scriptsize SemanticLock End points}}
		\label{fig:Web_study_password_EndPoint}
	\end{subfigure}
	\caption{{\footnotesize \textbf{Start/End points comparison}: Percentage representation of the Start and End points.}}
	\label{fig:StartEndpoints}
\end{figure*}
\subsection{Security \& Entropy Analysis}\label{Authen_Sec_Entropy}
With many authentication systems, users tend to choose passwords that are easy to remember, meaning that they do not select their password uniformly from the whole space of possible passwords, but instead show a higher probability to choose from certain subsets.  For example, PIN users often choose dates that have some significance to them as passwords.  The degree of randomness of passwords practically chosen by users is an important factor in determining the security of an authentication system. 


The primary attack we are considering is the brute force guessing attack. The objective of a guessing attack is to achieve a high number of match success within a fixed number of attempts, leveraging knowledge of user password preferences. Studies by \cite{Uellenbeck:2013:QSG:2508859.2516700,Andriotis:2013:PSS:2462096.2462098} proposed an algorithm called partial guessing entropy \cite{6234435} (\textit{$\alpha$-guesswork} ), which depicts the success rates as a function of the password distribution space. We use this algorithm to evaluate the security of SemanticLock with respect to guessing attacks.

In order to use $\alpha$-guesswork, we need to have an estimate of the distribution of user selected passwords. While the PIN password distribution can be estimated based on leaked password databases or surveys, it is more difficult to obtain this type of data for graphical password systems such as PATTERN and SemanticLock. Instead we use a Markov model from \cite{Uellenbeck:2013:QSG:2508859.2516700}, which is based on the idea that the subsequent token in a password, such as the next node in a PATTERN system, is dependent on the previous token. Therefore, with a given sequence of password tokens, we must determine the initial probabilities $P(c_{1},...,c_{m})$ and the subsequent transitional probabilities $P(c_{i}|c_{1},...,c_{n-1})$. This data was collected as part of our online survey.

\subsubsection{SemanticLock Web-based Data Analysis}
The data collected from the participants during stage 2 of the web-based study was analyzed to confirm that our icon selection method was valid and to derive statistics needed for our Markov model.

	\textbf{Password Icon distribution:} Frequency analysis was performed on the semantic password data sets collected. Each semantic password is made up of unique icons selected from the 6 initial password icons. From our dataset of 1825 semantically created passwords,  our analysis suggests that the choice of each of the six password icons is uniformly distributed.

	\textbf{Password Icon pair distribution:} As each semantic password is composed of two or more sets of password icons, we pre-processed the collected data sets and decomposed semantic passwords that consist of more than two password icons into two pairs of password icons and performed frequency analysis on these password icon pairs. All pairs were roughly equi-likely.

	\textbf{Password Icon-pair position distribution:} Password icons are used to create semantic passwords by dragging a selected password icon to  a \textit{``resting position''} next to the stationary password icon. This \textit{``resting position''} could either be the \textit{left, top, right} or \textit{bottom} of a stationary password icon (see Figure \ref{fig:web_9by6_page}(\subref{fig:pwd1})). We analyzed  the collected positional data sets to determine if our participants displayed a bias in their choice of \textit{``resting positions''}. Our analysis indicated that the participant selection of \textit{``resting positions``} was fairly uniform with a small bias towards the \textit{``top or right position``}, which is somewhat expected from predominantly right handed users.

\subsubsection{Password Strength Evaluation}
One objective for data collection during the online study was to quantify and compare the results obtained from the PATTERN and SemanticLock system. The metrics we obtained for pattern password evaluation were \textit{Pattern-length}, \textit{Stroke-length}, \textit{Intersections}, \textit{Start/End points} were similar to findings reported by \cite{Uellenbeck:2013:QSG:2508859.2516700,Andriotis:2013:PSS:2462096.2462098,Cha:2017:BGA:3052973.3052989,Harbach:2016:ASU:2858036.2858267}. 
The data collected and our analysis were highly similar to those reported in past studies by \cite{Aviv:2015:BBC:2818000.2818014,Cha:2017:BGA:3052973.3052989,Uellenbeck:2013:QSG:2508859.2516700,Bonneau:2012:SMI:2437647.2437657}. Implementing an accurate password strength comparison of the PATTERN and SemanticLock requires identifying metrics that are common to both systems or can be effectively generalized to serve our requirements. We determined that metrics such  \textit{Start/End} points and \textit{guess-ability resistance} are best suited for our comparison needs.
\begin{figure}[ht]
	\centering
	\includegraphics[width=0.7\columnwidth]{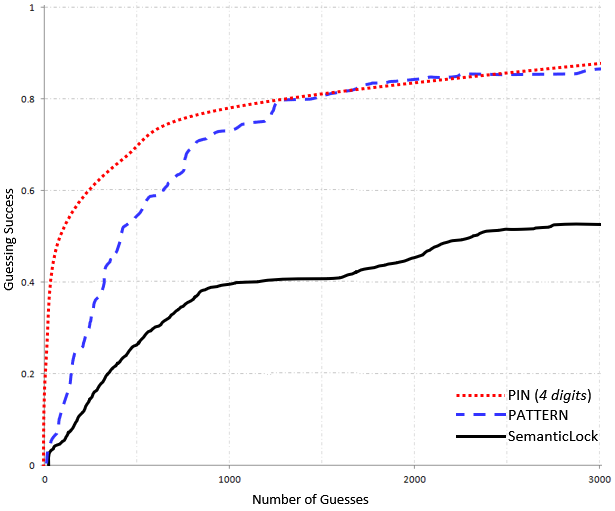}
\caption{{\footnotesize \textbf{Password Guessability Analysis : } Guessing entropy (\textit{$\alpha$-guesswork} ) comparison of the guessing resistance of Random PIN {\footnotesize (4 digits)}, PATTERN and SemanticLock. The graph of SemanticLock shows a high resistance to guessing attacks.}}
	\label{fig:guessgraph}
\end{figure}
			
\paragraph{\textbf{Password start and end Points:}}
The uniform distribution of \textit{start/end points} in a password system is an indication of high entropy and password strength \cite{Uellenbeck:2013:QSG:2508859.2516700,Andriotis:2013:PSS:2462096.2462098,Aviv:2015:BBC:2818000.2818014}. Analysis of the Pattern passwords collected during online study showed that 43.7\% of our participants started their password from the top-leftmost node, making their starting points highly predictable (see figure. \ref{fig:StartEndpoints}(\subref{fig:3_x_3_StartPoint})). Unsurprisingly, participants chose the bottom right node as their end destination 30.9\% of the time (see Figure. \ref{fig:StartEndpoints}(\subref{fig:3_x_3_endPoint})). These results are similar to findings observed by \cite{Uellenbeck:2013:QSG:2508859.2516700,Wiedenbeck:2005:PDL:1090412.1090418,Cha:2017:BGA:3052973.3052989}. Analysis of the SemanticLock passwords collected during online study shows the uniform distribution of start points (see figure . \ref{fig:StartEndpoints}(\subref{fig:Web_study_password_StartPointa})), with the largest value of 21.8\% located at the lower-rightmost cell, with a End point (see figure .\ref{fig:StartEndpoints}(\subref{fig:Web_study_password_EndPoint})) of 11.1\% located at the center of the grid. SemanticLock exhibited a lower level of bias and a more uniform distribution of participant password start and end points.



\paragraph{\textbf{Password Guessability:}}The results of our guessing attack evaluation is displayed in Figure \ref{fig:guessgraph}. In this figure, we depict guessing attack data for real user passwords, the PIN (4 digit) data was from a study by \cite{6234435}, and the PATTERN and SemanticLock data was collected during our web study. It can be seen that SemanticLock is more resistant to guessing attacks. For example, to compromise 20\%\textit{ (i.e $\alpha$ = 0.2)} of the password space of the PATTERN authentication system, it requires 114 attempts, while SemanticLock requires 346 attempts and PIN required less than 50 attempts. Additionally, to compromise 50\% \textit{(i.e $\alpha$ = 0.5)} of PATTERN, it requires 438 attempts, while SemanticLock requires 2422 attempts and PIN required less than 100 attempts.
\renewcommand{\arraystretch}{0.9} 
\begin{table}[h]
	\begin{center}
		\begin{tabular}{|l|c|c|c|} 
			\hline  
			\rule{0pt}{2pt} {\scriptsize Distribution}	& {\scriptsize $\alpha = 0.1$} & {\scriptsize $\alpha = 0.2$} & {\scriptsize $\alpha = 0.5$}\\	
			\hline
			{\scriptsize \textbf{SemanticLock}} & {\scriptsize \textbf{9.89}} & {\scriptsize \textbf{10.26}}  & {\scriptsize \textbf{11.7}} \\		
			{\scriptsize PATTERN 3x3  } &	{\scriptsize 7.10}	& {\scriptsize 7.86}  & {\scriptsize 9.98}  \\
			\hline
		{\scriptsize RealUser PIN \textit{(4 digits)} }{\tiny \cite{Kim:2012:PSP:2622683.2623002,10.1007/978-3-642-32946-3_3,6234435}} &	{\scriptsize 5.19}	& {\scriptsize 7.04}  & {\scriptsize 10.08}  \\	
		\hline
			{\scriptsize PATTERN 3x3 } {\tiny (Tupsamudre et.al) \cite{Tupsamudre:2017:PPI:3052973.3053041}} &	{\scriptsize 5.80}	& {\scriptsize 6.95}  & {\scriptsize 9.86}  \\	
			{\scriptsize PATTERN 3x3 } {\tiny (Aviv et.al) \cite{Aviv:2015:BBC:2818000.2818014}} &	{\scriptsize 6.59}	& {\scriptsize 6.99}  & {\scriptsize 8.93}  \\
			{\scriptsize PATTERN 3x3 } {\tiny (Uellenbeck et.al) \cite{Uellenbeck:2013:QSG:2508859.2516700}} &	{\scriptsize 8.72}	& {\scriptsize 9.10}  & {\scriptsize 10.90}  \\
			\hline
		\end{tabular}
		\caption{{\footnotesize \textbf{Partial Guessing Entropy Comparison}: This chart compares partial entropy estimates of several distributions and different values for the   (\textit{$\alpha$-guesswork} ) }}
\label{tab:PartialGuess}	
	\end{center}
\end{table}
\renewcommand{\arraystretch}{1} 

Our results are shown in Table \ref{tab:PartialGuess} along with partial entropy estimates from other studies. We computed entropy estimates for $\alpha$=10\%, 20\% and 50\%, higher values of $\alpha$ for non-uniform distributions reflect a higher entropy factor. From Table \ref{tab:PartialGuess}, we note that SemanticLock has a better performance factor than all the \textit{``practical" }PATTERN$_{(Tupsamudre,Aviv,Uellenbeck,Olade)}$ and RealUser PIN( 4 digit) estimates, with its $\alpha$ values significantly higher than the password strength of a uniformly distributed 3-digit Random PIN.

\subsection{Quantitative Results} \label{Quantitative_Results}
\subsubsection{Login Speed}
\begin{figure}[h]
		\centering
		\includegraphics[height=\imgHeightRatio\paperheight]{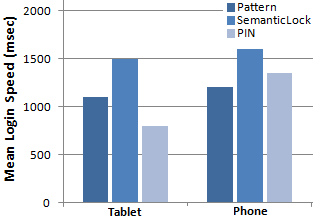}
		\caption{ {\footnotesize \textbf{Mean Login Speed: } Login speed while seated with different device form.}}
		\label{fig:Mean_Login_Speed_Tablet_Phone1}
\end{figure}

The mean values of the login speed of each technique and other independent factors are shown in Figure \ref{fig:Mean_Login_Speed_Tablet_Phone1}. The results show that the SemanticLock performed better than the other techniques across device form factors. SemanticLock was superior in performance to PIN across all independent variables. There was a statistically significant difference between the techniques login speed as determined by the one-way ANOVA test (F(4,535) = 170.44, p $<$ .001). A Tukey post hoc test revealed that SemanticLock (807.06 $\pm$ 167.23 ms, p $<$ .001) was significantly faster than Pattern and PIN (both p $<$ .001). 


\subsubsection{Differences across Device Form Factors}
As stated earlier, we used two different types of device form-factors during the sessions (a Nexus 5 phone and a Google Pixel C tablet (see Figure. \ref{fig:Mean_Login_Speed_Tablet_Phone1}). Results of a two-way ANOVA test show that there was no significant effect of device form-factor ( F(1,530) = .003, p = .995) on login speed across techniques. Furthermore there was no significant interaction effect between device form-factor and login technique (F(4,530) = 1.208, p = .306),  (see figure. \ref{fig:Mean_Login_Speed_Tablet_Phone1}).

\subsubsection{Pre-Login Delay Time}
Our participants experience a time delay between when the trial started and when an initial action or interaction was made. This pre-login delay time gives an indication of familiarity, memorability or ease of use of the techniques. SemanticLock had the lowest pre-login delay time, the ANOVA test results showed a significant main effect for hand input posture, (F(2,930) = 9.877, p $<$ 0.05).
 
\subsubsection{Error Rate}

\begin{figure}[ht]
	\centering
	\includegraphics[width=0.70\columnwidth,height=\imgHeightRatio\paperheight]{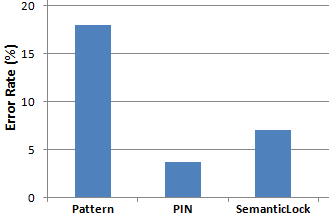}
	\caption{\footnotesize{Error rates for each Technique}}
	\label{fig:Error_rate_for_each_Techniques}
\end{figure}

A two-way ANOVA test  was conducted to examine the error rate for each technique. There was no significant effect of interaction by these independent variables on the error rate. Furthermore, analysis showed that error rate was lowest for SemanticLock and there was no significant difference in the error rate of the SemanticLock technique (p = .925 ). Additionally, results shows that error rates classified by techniques show that Pattern (18\%) had the highest error rates, followed by  SemanticLock (7\%), and PIN(3.5\%)  (see figure. \ref{fig:Error_rate_for_each_Techniques}).

\subsection{Qualitative Results}
\begin{figure}[h]
	\centering
	\includegraphics[width=0.70\columnwidth,height=\imgHeightRatio\paperheight]{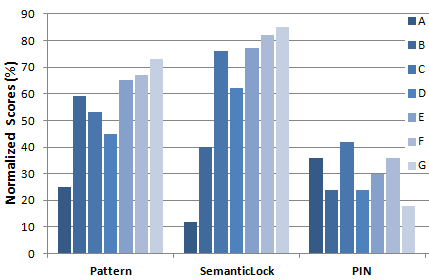}
	\caption{{\footnotesize \textbf{User LIKERT ranking survey}: Our LIKERT based qualitative test indicates that the SemantickLock performed better with all the evaluated factors \textit{(see legend A to G)}. \textit{[ A: Hard to Recall, B: Best GUI , C: Easy to Recall , D: Use In Future , E: Liked the Most , F: Easy to Use , G: Faster Login ]}}}
	\label{fig:Post_Test_user_ranking_survey}
\end{figure}

The results are based on a 5-point Likert scale questionnaire and subsequent user rankings of the three techniques. Each participant prior to the experiment answered an electronic pre-test survey which we used to obtain demographics, personal information, and mobile device usage experience. The Likert scaled questions were answered after the trial of each technique to collect their subjective preferences. At the end a user ranking of all techniques was collected  (see Figure. \ref{fig:Post_Test_user_ranking_survey}). The data we collected was analyzed using the Friedman test and we performed post hoc analysis with Wilcoxon signed-rank test with Bonferroni correction (p= 0.05/3 = 0.017) of those that are statistically significant. In the questionnaire we probed aspects of the users experience with the three login techniques.
\begin{figure}[H]
 	\centering
 	\includegraphics[width=0.70\columnwidth]{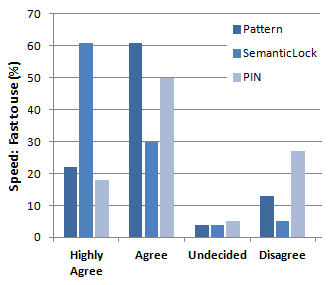}
 	\caption{\footnotesize{\textbf{Perceived Login Speed: }A comparison of the users' perceived login speed for each technique.}}
 	\label{fig:Likert_Speed_Fast}
 \end{figure}
\subsubsection{Speed}
Our participants experience with each technique's speed shows there was a statistically significant perceived difference in speed depending on technique ($\chi^{2(2)}$ = 18.321, p $<$ 0.001) (see Figure.\ref{fig:Likert_Speed_Fast}). Post hoc analysis indicated that there were no significant differences between PIN and Pattern trials (Z = -2.101, p = 0.036) or between PIN and SemanticLock trials (Z = -1.560, p = 0.119). However, there was significant difference in speed between Pattern and SemanticLock trials (Z = -3.573, p $<$ 0.001). 
 
\begin{figure*}[t]
		\centering
		\begin{subfigure}[b]{.47\textwidth}
		\centering
		\includegraphics[height=\imgHeightRatio\paperheight]{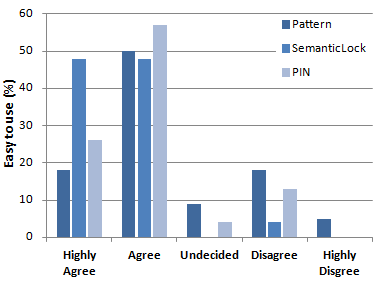}
		\caption{Easy to use}
		\label{fig:Likert_Easy_to_use}
	\end{subfigure}\hfill
	\begin{subfigure}[b]{.47\textwidth}
		\centering
		\includegraphics[height=\imgHeightRatio\paperheight]{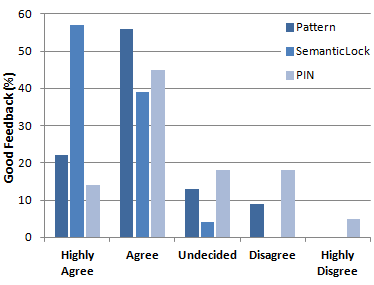}
		\caption{Positive Opinion}
		\label{fig:Likert_GoodFeedback}
	\end{subfigure}
	\caption{{\footnotesize \textbf{Qualitative Results}: Evaluation of the qualitative results using the LIKERT scale indicates that  \textbf{(a)} Results also indicates that 48\% of participants believe that SemanticLock was ``easy to use''. \textbf{(b)}  57\% had a positive opinion of SemanticLock.}}	
\end{figure*}
\subsubsection{Likeability}
Post hoc analysis indicated that there was no significant difference in how well participants liked the techniques (see figure .\ref{fig:Post_Test_user_ranking_survey}).

\subsubsection{Usability}
There was a significant difference in perceived ease of use of technique ($\chi^{2(2)}$ = 14.22, p = 0.001). Post hoc analysis indicated that there were no significant differences between the PIN and Pattern (Z = -1.672, p = 0.94) or between the PIN and SemanticLock (Z = -1.628, p = 0.103)  (see figure. \ref{fig:Likert_Easy_to_use}). However, there was a significant increase in perceived ease of use between Pattern and SemanticLock (Z = -3.140, p = 0.002). 

\subsubsection{Positive Feedback}
Participants experience with the feedback for each technique also showed that there was a significant difference ($\chi^{2(2)}$ = 17.179, p $<$ 0.001)  (see figure.\ref{fig:Likert_GoodFeedback}). There were significant differences between Pattern and SemanticLock as well as SemanticLock and PIN; SemanticLock was ranked favorably in both cases.
\subsubsection{Error Recovery}
There was a significant difference in error recovery based on technique ($\chi^{2(2)}$ = 12.667, p = 0.002). Significant differences were found between Pattern and SemanticLock as well as PIN and SemanticLock. In both cases, Pattern and PIN were ranked favorably in regards to ease of error recovery. There was no significant difference in how participants liked interacting with the techniques.
\begin{figure}[h]
	\centering
	\includegraphics[width=0.90\columnwidth]{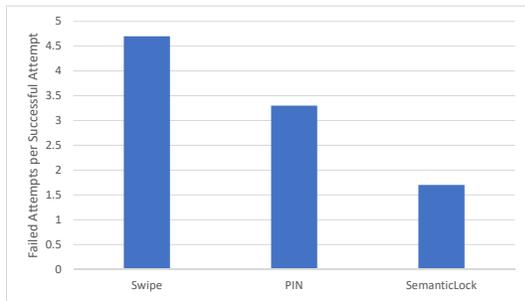}
	\caption{{\footnotesize Correct vs incorrect guesses after 3 weeks.}}
	\label{fig:ErrorRateWeek3}
\end{figure}
\subsection{Memorability}
We split our memorability results into two parts.  Figure \ref{fig:ErrorRateWeek3} shows the average number of unsuccessful attempts before successfully entering the correct password for the various methods.  In this figure we see that Swipe was the most difficult to enter correctly, followed by PIN, and then SemanticLock.

In Figure \ref{fig:ErrorRateWeek3} we can see the failure rate for the three techniques.  Failure is defined as the complete inability to remember the password.  Again, Swipe fared the worst of the three methods, followed by PIN, then SemanticLock. These results support feedback from users, who felt the abstract nature of the randomized Swipe patterns were very hard to remember.  PINs were easier to remember, but prone to single digit errors, or digit swapping errors.  

\subsection{Long-term Usability}
At the end of the two week long-term usability experiment we gave participants a short Likert-scale based survey and discussed their SemanticLock experience for anecdotal feedback.  None of the users reported any issues with the SemanticLock software, and all users were still using the software at the end of the two week trial.  Questions on the survey included:
\begin{enumerate}
\item Did you enjoy using SemanticLock (1 - not at all to 5 - very much) \textbf{(Average: 4.1)}
\item Do you feel that SemanticLock could replace the usual phone unlocking method? (1 - strongly disagree to 5 - strongly agree) \textbf{(Average: 4.2)}
\item Do you feel that SemanticLock was slower to use than your usual method (1 - much slower to 5 - much faster) \textbf{(Average: 3.1)}
\item How easy was your SemanticLock password to remember? (1 - very difficult to 5 very easy) \textbf{(Average: 4.6)}
\item Did you make more or fewer errors with SemanticLock than your usual phone unlocking method? (1 - a lot more to 5 - a lot fewer) \textbf{(Average: 3.2)}
\item What is your opinion of the SemanticLock? (1 - very unfavorable to 5 very favorable) \textbf{(Average: 4.5)}
\end{enumerate}

From these survey results we can see that after two weeks of constant use, participants felt that SemanticLock was as usable as their existing authentication method, and enjoyed using it.  User perception of speed and error rates were that they were roughly equivalent to PIN or Swipe.  Feedback from the post experiment interview generally focused around design details and enhancements, for example several users thought the icon sets were a bit dull and could be more colorful.  Other users requested a choice of icon sets with different themes or styles, or the ability to customize the background or size of the icons.
\begin{figure}[h]
	\centering
	\includegraphics[width=0.90\columnwidth]{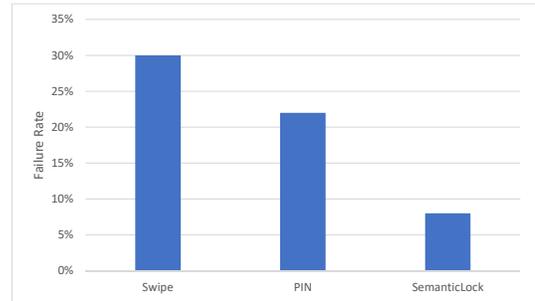}
	\caption{{\footnotesize Failure rates for users after 3 weeks.}}
	\label{fig:FailureRateWeek3}
\end{figure}
\section{Study Limitations and Future Work} \label{Study_limits}
Although our study had a relatively long duration, we were not able to evaluate the very long-term memorability effects of SemanticLock. We believe that  SemanticLock performance will benefit when users have more practice and familiarity with it. In regards to generalization, our sample population represents the most common users of mobile devices and should be able to generalize to other populations. We believe that other populations like children and the elderly will be inclined to use SemanticLock, and we feel this would be a useful area to explore.  We also feel that while the idea of designing the password space to encourage high entropy user selected passwords is quite powerful, there is further work to be done.  For example, our participant groups were all of a similar cultural and educational background.  It is probable that other groups of users would have different sets of biases, requiring customization of the icon sets based on region or age group.

\section{Conclusion} \label{Discussion}
Our design strategy for SemanticLock was to create an authentication method that was as usable as conventional methods, while offering better memorability and security.  Our experimental data shows that SemanticLock has achieved exactly that by combining semantically meaningful story based password to improve memorability with a carefully designed password space to improve user selected password entropy.

Data analysis indicates that SemanticLock clearly has a stronger practical password strength than the PATTERN or PIN authentication system. Results from section \ref{Authen_Sec_Entropy} shows that SemanticLock has little or no password start/end point bias (see Figure \ref{fig:StartEndpoints}). Furthermore, evaluations performed using partial guessing entropy shows that the practical entropy of SemanticLock is closer to the security offered by a uniformly distributed Random 4-digit PIN and outperformed all the practical strength of the PATTERN authentication system examined in this paper (see Table \ref{tab:PartialGuess}).

Our participants both quantitatively and qualitatively found SemanticLock as usable as current mainstream authentication methods. Our memorability study showed that users retained SemanticLock passwords much more easily that PIN or Swipe, even after two weeks of non-use.  This is a significant advantage that is particularly applicable for devices that aren't regularly used, or for cases where traditional authentication is a backup method, for example with biometric authentication. Our long-term study users found SemanticLock both enjoyable to use and a viable alternative to either PIN or Swipe as a primary authentication mechanism.

\begin{figure*}[H]
		\centering
		\begin{subfigure}[b]{.47\textwidth}
        	\centering
        	\includegraphics[height=\imgHeightRatio\paperheight]{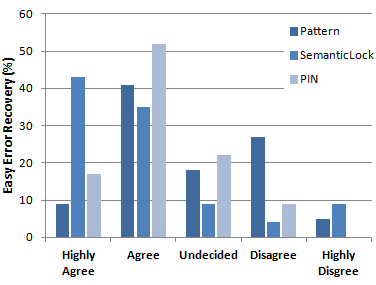}
        	\caption{Error Recovery}
        	\label{fig:Likert_Easy_error_recovery}
	\end{subfigure}\hfill
	\begin{subfigure}[b]{.47\textwidth}
		\centering
		\includegraphics[height=\imgHeightRatio\paperheight]{images/Likert_Speed_Fast}
		\caption{Fast Login Speed }
		\label{fig:Likert_Speed_Fast}
	\end{subfigure}
	\caption{{\footnotesize \textbf{Qualitative Results}: Evaluation of the qualitative results using the LIKERT scale indicates that  \textbf{(a)} Results also indicates that 43\% observed easy error recovery when using SemanticLock. \textbf{(b)} 60\% of participants observed that SemanticLock has a faster Login speed.}}		
\end{figure*}




{\footnotesize \bibliographystyle{acm}
\bibliography{semanticlockRef}}

\begin{thebibliography}{10}

\bibitem{Abdelrahman:2017:SCU:3025453.3025461}
{\sc Abdelrahman, Y., Khamis, M., Schneegass, S., and Alt, F.}
\newblock Stay cool! understanding thermal attacks on mobile-based user
  authentication.
\newblock In {\em Proceedings of the 2017 CHI Conference on Human Factors in
  Computing Systems\/} (New York, NY, USA, 2017), CHI '17, ACM, pp.~3751--3763.

\bibitem{Aly:2016:SGA:2957265.2961863}
{\sc Aly, Y., Munteanu, C., Raimondo, S., Wu, A.~Y., and Wei, M.}
\newblock Spin-lock gesture authentication for mobile devices.
\newblock In {\em Proceedings of the 18th International Conference on
  Human-Computer Interaction with Mobile Devices and Services Adjunct\/} (New
  York, NY, USA, 2016), MobileHCI '16, ACM, pp.~775--782.

\bibitem{Andriotis:2013:PSS:2462096.2462098}
{\sc Andriotis, P., Tryfonas, T., Oikonomou, G., and Yildiz, C.}
\newblock A pilot study on the security of pattern screen-lock methods and soft
  side channel attacks.
\newblock In {\em Proceedings of the Sixth ACM Conference on Security and
  Privacy in Wireless and Mobile Networks\/} (New York, NY, USA, 2013), WiSec
  '13, ACM, pp.~1--6.

\bibitem{Aviv:2015:BBC:2818000.2818014}
{\sc Aviv, A.~J., Budzitowski, D., and Kuber, R.}
\newblock Is bigger better? comparing user-generated passwords on 3x3 vs. 4x4
  grid sizes for android's pattern unlock.
\newblock In {\em Proceedings of the 31st Annual Computer Security Applications
  Conference\/} (New York, NY, USA, 2015), ACSAC 2015, ACM, pp.~301--310.

\bibitem{Belk:2017:SSU:3106426.3106488}
{\sc Belk, M., Pamboris, A., Fidas, C., Katsini, C., Avouris, N., and Samaras,
  G.}
\newblock Sweet-spotting security and usability for intelligent graphical
  authentication mechanisms.
\newblock In {\em Proceedings of the International Conference on Web
  Intelligence\/} (New York, NY, USA, 2017), WI '17, ACM, pp.~252--259.

\bibitem{Biddle:2012:GPL:2333112.2333114}
{\sc Biddle, R., Chiasson, S., and Van~Oorschot, P.}
\newblock Graphical passwords: Learning from the first twelve years.
\newblock {\em ACM Comput. Surv. 44}, 4 (Sept. 2012), 19:1--19:41.

\bibitem{BlonderG69}
{\sc Blonder, G.}
\newblock Graphical password.
\newblock {\em In Lucent Technologies, Inc\/} (1996.).

\bibitem{6234435}
{\sc Bonneau, J.}
\newblock The science of guessing: Analyzing an anonymized corpus of 70 million
  passwords.
\newblock In {\em 2012 IEEE Symposium on Security and Privacy\/} (May 2012),
  pp.~538--552.

\bibitem{Bonneau:2012:SMI:2437647.2437657}
{\sc Bonneau, J.}
\newblock Statistical metrics for individual password strength.
\newblock In {\em Proceedings of the 20th International Conference on Security
  Protocols\/} (Berlin, Heidelberg, 2012), SP'12, Springer-Verlag, pp.~76--86.

\bibitem{Bonneau:2015:PEI:2797100.2699390}
{\sc Bonneau, J., Herley, C., van Oorschot, P.~C., and Stajano, F.}
\newblock Passwords and the evolution of imperfect authentication.
\newblock {\em Commun. ACM 58}, 7 (June 2015), 78--87.

\bibitem{10.1007/978-3-642-32946-3_3}
{\sc Bonneau, J., Preibusch, S., and Anderson, R.}
\newblock A birthday present every eleven wallets? the security of
  customer-chosen banking pins.
\newblock In {\em Financial Cryptography and Data Security\/} (Berlin,
  Heidelberg, 2012), A.~D. Keromytis, Ed., Springer Berlin Heidelberg,
  pp.~25--40.

\bibitem{10.1007/978-1-4471-0515-2_27}
{\sc Brostoff, S., and Sasse, M.~A.}
\newblock Are passfaces more usable than passwords? a field trial
  investigation.
\newblock In {\em People and Computers XIV --- Usability or Else!\/} (London,
  2000), S.~McDonald, Y.~Waern, and G.~Cockton, Eds., Springer London,
  pp.~405--424.

\bibitem{buschek_snapapp_2016}
{\sc Buschek, D., Hartmann, F., von Zezschwitz, E., De~Luca, A., and Alt, F.}
\newblock {SnapApp}: {Reducing} {Authentication} {Overhead} with a
  {Time}-{Constrained} {Fast} {Unlock} {Option}.
\newblock In {\em Proceedings of the 2016 {CHI} {Conference} on {Human}
  {Factors} in {Computing} {Systems}\/} (New York, NY, USA, 2016), {CHI} '16,
  ACM, pp.~3736--3747.

\bibitem{Cain:2017:GAR:3027063.3053236}
{\sc Cain, A.~A., Werner, S., and Still, J.~D.}
\newblock Graphical authentication resistance to over-the-shoulder-attacks.
\newblock In {\em Proceedings of the 2017 CHI Conference Extended Abstracts on
  Human Factors in Computing Systems\/} (New York, NY, USA, 2017), CHI EA '17,
  ACM, pp.~2416--2422.

\bibitem{Cha:2017:BGA:3052973.3052989}
{\sc Cha, S., Kwag, S., Kim, H., and Huh, J.~H.}
\newblock Boosting the guessing attack performance on android lock patterns
  with smudge attacks.
\newblock In {\em Proceedings of the 2017 ACM on Asia Conference on Computer
  and Communications Security\/} (New York, NY, USA, 2017), ASIA CCS '17, ACM,
  pp.~313--326.

\bibitem{Chiang:2013:IUA:2493190.2493213}
{\sc Chiang, H.-Y., and Chiasson, S.}
\newblock Improving user authentication on mobile devices: A touchscreen
  graphical password.
\newblock In {\em Proceedings of the 15th International Conference on
  Human-computer Interaction with Mobile Devices and Services\/} (New York, NY,
  USA, 2013), MobileHCI '13, ACM, pp.~251--260.

\bibitem{Chiasson:2009:UID:1667545.1667546}
{\sc Chiasson, S., Forget, A., Biddle, R., and van Oorschot, P.~C.}
\newblock User interface design affects security: Patterns in click-based
  graphical passwords.
\newblock {\em Int. J. Inf. Secur. 8}, 6 (Oct. 2009), 387--398.

\bibitem{Chiasson:2007:GPA:2393847.2393880}
{\sc Chiasson, S., Van~Oorschot, P.~C., and Biddle, R.}
\newblock Graphical password authentication using cued click points.
\newblock In {\em Proceedings of the 12th European Conference on Research in
  Computer Security\/} (Berlin, Heidelberg, 2007), ESORICS'07, Springer-Verlag,
  pp.~359--374.

\bibitem{Davis:2004:UCG:1251375.1251386}
{\sc Davis, D., Monrose, F., and Reiter, M.~K.}
\newblock On user choice in graphical password schemes.
\newblock In {\em Proceedings of the 13th Conference on USENIX Security
  Symposium - Volume 13\/} (Berkeley, CA, USA, 2004), SSYM'04, USENIX
  Association, pp.~11--11.

\bibitem{DeLuca:2010:CSP:1753326.1753490}
{\sc De~Luca, A., Hertzschuch, K., and Hussmann, H.}
\newblock Colorpin: Securing pin entry through indirect input.
\newblock In {\em Proceedings of the SIGCHI Conference on Human Factors in
  Computing Systems\/} (New York, NY, USA, 2010), CHI '10, ACM, pp.~1103--1106.

\bibitem{Dhamija:2000:DVU:1251306.1251310}
{\sc Dhamija, R., and Perrig, A.}
\newblock D{\'e}j\`{a} vu: A user study using images for authentication.
\newblock In {\em Proceedings of the 9th Conference on USENIX Security
  Symposium - Volume 9\/} (Berkeley, CA, USA, 2000), SSYM'00, USENIX
  Association, pp.~4--4.

\bibitem{Dunphy:2010:CLR:1837110.1837114}
{\sc Dunphy, P., Heiner, A.~P., and Asokan, N.}
\newblock A closer look at recognition-based graphical passwords on mobile
  devices.
\newblock In {\em Proceedings of the Sixth Symposium on Usable Privacy and
  Security\/} (New York, NY, USA, 2010), SOUPS '10, ACM, pp.~3:1--3:12.

\bibitem{Fahl:2013:EVP:2501604.2501617}
{\sc Fahl, S., Harbach, M., Acar, Y., and Smith, M.}
\newblock On the ecological validity of a password study.
\newblock In {\em Proceedings of the Ninth Symposium on Usable Privacy and
  Security\/} (New York, NY, USA, 2013), SOUPS '13, ACM, pp.~13:1--13:13.

\bibitem{Feng:2015:IPI:2786567.2793711}
{\sc Feng, S., Wilson, G., Ng, A., and Brewster, S.}
\newblock Investigating pressure-based interactions with mobile phones while
  walking and encumbered.
\newblock In {\em Proceedings of the 17th International Conference on
  Human-Computer Interaction with Mobile Devices and Services Adjunct\/} (New
  York, NY, USA, 2015), MobileHCI '15, ACM, pp.~854--861.

\bibitem{Haque:2013:PIT:2516760.2516767}
{\sc Haque, S. M.~T., Wright, M., and Scielzo, S.}
\newblock Passwords and interfaces: Towards creating stronger passwords by
  using mobile phone handsets.
\newblock In {\em Proceedings of the Third ACM Workshop on Security and Privacy
  in Smartphones \&\#38; Mobile Devices\/} (New York, NY, USA, 2013), SPSM '13,
  ACM, pp.~105--110.

\bibitem{Harbach:2016:ASU:2858036.2858267}
{\sc Harbach, M., De~Luca, A., and Egelman, S.}
\newblock The anatomy of smartphone unlocking: A field study of android lock
  screens.
\newblock In {\em Proceedings of the 2016 CHI Conference on Human Factors in
  Computing Systems\/} (New York, NY, USA, 2016), CHI '16, ACM, pp.~4806--4817.

\bibitem{Harbach}
{\sc Harbach, M., von Zezschwitz, E., Fichtner, A., Luca, A.~D., and Smith, M.}
\newblock It{\textquoteright}s a hard lock life: A field study of smartphone
  (un)locking behavior and risk perception.
\newblock In {\em 10th Symposium On Usable Privacy and Security ({SOUPS}
  2014)\/} (Menlo Park, CA, 2014), {USENIX} Association, pp.~213--230.

\bibitem{Jakobsson:2012:BUP:2372387.2372397}
{\sc Jakobsson, M., and Dhiman, M.}
\newblock The benefits of understanding passwords.
\newblock In {\em Proceedings of the 7th USENIX Conference on Hot Topics in
  Security\/} (Berkeley, CA, USA, 2012), HotSec'12, USENIX Association,
  pp.~10--10.

\bibitem{Kim:2012:PSP:2622683.2623002}
{\sc Kim, H., and Huh, J.~H.}
\newblock Pin selection policies: Are they really effective?
\newblock {\em Comput. Secur. 31}, 4 (June 2012), 484--496.

\bibitem{kovelamudi_scramble_2016}
{\sc Kovelamudi, G., Zheng, J., and Mukkamala, S.}
\newblock Scramble or not, that is the question a study of the security and
  usability of scramble keypad for {PIN} unlock on smartphones.
\newblock In {\em 2016 {IEEE}/{CIC} {International} {Conference} on
  {Communications} in {China} ({ICCC})\/} (July 2016), pp.~1--6.

\bibitem{Malone:2012:IDP:2187836.2187878}
{\sc Malone, D., and Maher, K.}
\newblock Investigating the distribution of password choices.
\newblock In {\em Proceedings of the 21st International Conference on World
  Wide Web\/} (New York, NY, USA, 2012), WWW '12, ACM, pp.~301--310.

\bibitem{Melicher:2016:UST:2858036.2858384}
{\sc Melicher, W., Kurilova, D., Segreti, S.~M., Kalvani, P., Shay, R., Ur, B.,
  Bauer, L., Christin, N., Cranor, L.~F., and Mazurek, M.~L.}
\newblock Usability and security of text passwords on mobile devices.
\newblock In {\em Proceedings of the 2016 CHI Conference on Human Factors in
  Computing Systems\/} (New York, NY, USA, 2016), CHI '16, ACM, pp.~527--539.

\bibitem{Micallef:2015:WAU:2785830.2785835}
{\sc Micallef, N., Just, M., Baillie, L., Halvey, M., and Kayacik, H.~G.}
\newblock Why aren't users using protection? investigating the usability of
  smartphone locking.
\newblock In {\em Proceedings of the 17th International Conference on
  Human-Computer Interaction with Mobile Devices and Services\/} (New York, NY,
  USA, 2015), MobileHCI '15, ACM, pp.~284--294.

\bibitem{Mowery:2011:HMC:2028052.2028058}
{\sc Mowery, K., Meiklejohn, S., and Savage, S.}
\newblock Heat of the moment: Characterizing the efficacy of thermal
  camera-based attacks.
\newblock In {\em Proceedings of the 5th USENIX Conference on Offensive
  Technologies\/} (Berkeley, CA, USA, 2011), WOOT'11, USENIX Association,
  pp.~6--6.

\bibitem{Ng:2014:IEE:2556288.2557312}
{\sc Ng, A., Brewster, S.~A., and Williamson, J.~H.}
\newblock Investigating the effects of encumbrance on one- and two- handed
  interactions with mobile devices.
\newblock In {\em Proceedings of the SIGCHI Conference on Human Factors in
  Computing Systems\/} (New York, NY, USA, 2014), CHI '14, ACM, pp.~1981--1990.

\bibitem{Ng:2015:EEM:2785830.2785853}
{\sc Ng, A., Williamson, J., and Brewster, S.}
\newblock The effects of encumbrance and mobility on touch-based gesture
  interactions for mobile phones.
\newblock In {\em Proceedings of the 17th International Conference on
  Human-Computer Interaction with Mobile Devices and Services\/} (New York, NY,
  USA, 2015), MobileHCI '15, ACM, pp.~536--546.

\bibitem{Passfaces1}
{\sc {Passfaces}}.
\newblock Passfaces: Two factor authentication for the enterprise, 2018.
\newblock [Online; accessed March 27, 2018].

\bibitem{Riva:2012:PAD:2362793.2362808}
{\sc Riva, O., Qin, C., Strauss, K., and Lymberopoulos, D.}
\newblock Progressive authentication: Deciding when to authenticate on mobile
  phones.
\newblock In {\em Proceedings of the 21st USENIX Conference on Security
  Symposium\/} (Berkeley, CA, USA, 2012), Security'12, USENIX Association,
  pp.~15--15.

\bibitem{Stobert:2013:MRG:2501604.2501619}
{\sc Stobert, E., and Biddle, R.}
\newblock Memory retrieval and graphical passwords.
\newblock In {\em Proceedings of the Ninth Symposium on Usable Privacy and
  Security\/} (New York, NY, USA, 2013), SOUPS '13, ACM, pp.~15:1--15:14.

\bibitem{Tao2008PassGoAP}
{\sc Tao, H., and Adams, C.~M.}
\newblock Pass-go: A proposal to improve the usability of graphical passwords.
\newblock {\em I. J. Network Security 7\/} (2008), 273--292.

\bibitem{Tupsamudre:2017:PPI:3052973.3053041}
{\sc Tupsamudre, H., Banahatti, V., Lodha, S., and Vyas, K.}
\newblock Pass-o: A proposal to improve the security of pattern unlock scheme.
\newblock In {\em Proceedings of the 2017 ACM on Asia Conference on Computer
  and Communications Security\/} (New York, NY, USA, 2017), ASIA CCS '17, ACM,
  pp.~400--407.

\bibitem{Uellenbeck:2013:QSG:2508859.2516700}
{\sc Uellenbeck, S., D\"{u}rmuth, M., Wolf, C., and Holz, T.}
\newblock Quantifying the security of graphical passwords: The case of android
  unlock patterns.
\newblock In {\em Proceedings of the 2013 ACM SIGSAC Conference on Computer
  \&\#38; Communications Security\/} (New York, NY, USA, 2013), CCS '13, ACM,
  pp.~161--172.

\bibitem{Veras:2012:VSP:2379690.2379702}
{\sc Veras, R., Thorpe, J., and Collins, C.}
\newblock Visualizing semantics in passwords: The role of dates.
\newblock In {\em Proceedings of the Ninth International Symposium on
  Visualization for Cyber Security\/} (New York, NY, USA, 2012), VizSec '12,
  ACM, pp.~88--95.

\bibitem{vonZezschwitz:2015:SFS:2702123.2702212}
{\sc von Zezschwitz, E., De~Luca, A., Brunkow, B., and Hussmann, H.}
\newblock Swipin: Fast and secure pin-entry on smartphones.
\newblock In {\em Proceedings of the 33rd Annual ACM Conference on Human
  Factors in Computing Systems\/} (New York, NY, USA, 2015), CHI '15, ACM,
  pp.~1403--1406.

\bibitem{vonZezschwitz:2013:PWF:2493190.2493231}
{\sc von Zezschwitz, E., Dunphy, P., and De~Luca, A.}
\newblock Patterns in the wild: A field study of the usability of pattern and
  pin-based authentication on mobile devices.
\newblock In {\em Proceedings of the 15th International Conference on
  Human-computer Interaction with Mobile Devices and Services\/} (New York, NY,
  USA, 2013), MobileHCI '13, ACM, pp.~261--270.

\bibitem{vonZezschwitz:2013:MGA:2449396.2449432}
{\sc von Zezschwitz, E., Koslow, A., De~Luca, A., and Hussmann, H.}
\newblock Making graphic-based authentication secure against smudge attacks.
\newblock In {\em Proceedings of the 2013 International Conference on
  Intelligent User Interfaces\/} (New York, NY, USA, 2013), IUI '13, ACM,
  pp.~277--286.

\bibitem{Weiss:2008:PUS:1463160.1463202}
{\sc Weiss, R., and De~Luca, A.}
\newblock Passshapes: Utilizing stroke based authentication to increase
  password memorability.
\newblock In {\em Proceedings of the 5th Nordic Conference on Human-computer
  Interaction: Building Bridges\/} (New York, NY, USA, 2008), NordiCHI '08,
  ACM, pp.~383--392.

\bibitem{WIEDENBECK2005102}
{\sc Wiedenbeck, S., Waters, J., Birget, J.-C., Brodskiy, A., and Memon, N.}
\newblock Passpoints: Design and longitudinal evaluation of a graphical
  password system.
\newblock {\em International Journal of Human-Computer Studies 63}, 1 (2005),
  102 -- 127.
\newblock HCI research in privacy and security.

\bibitem{Wiedenbeck:2005:PDL:1090412.1090418}
{\sc Wiedenbeck, S., Waters, J., Birget, J.-C., Brodskiy, A., and Memon, N.}
\newblock Passpoints: Design and longitudinal evaluation of a graphical
  password system.
\newblock {\em Int. J. Hum.-Comput. Stud. 63}, 1-2 (July 2005), 102--127.

\bibitem{Zakaria:2011:SSD:2078827.2078835}
{\sc Zakaria, N.~H., Griffiths, D., Brostoff, S., and Yan, J.}
\newblock Shoulder surfing defence for recall-based graphical passwords.
\newblock In {\em Proceedings of the Seventh Symposium on Usable Privacy and
  Security\/} (New York, NY, USA, 2011), SOUPS '11, ACM, pp.~6:1--6:12.

\end{thebibliography}

\end{document}